\documentclass[11pt]{article}

\newcommand{\be}{\begin{equation}}
\newcommand{\ee}{\end{equation}}
\newcommand{\bea}{\begin{eqnarray}}
\newcommand{\eea}{\end{eqnarray}}

\DeclareMathSymbol{\mg}{\mathrel}{symbols}{"1D}

\newcommand{\ga}{\alpha}

\renewcommand{\ge}{\epsilon}

\newcommand{\gth}{\theta}

\newcommand{\gp}{\pi}

\newcommand{\get}{\eta}

\newcommand{\cJ}{{\cal J}}

\newcommand{\cN}{{\cal N}}
\newcommand{\cO}{{\cal O}}

\newcommand{\cR}{{\cal R}}

\newcommand{\bZ}{{\bf Z}}
\newcommand{\ra}{\rightarrow}

\newcounter{oldcounter}

\begin{document}
\begin{flushright}
hep-ph/0212197,\\ 
DAMTP-2002-160, \\
UVIC-TH/02-05.
\end{flushright} 
\vskip 1.5cm
\begin{center} 
{\Large {\bf  String corrections to  gauge couplings \\
\bigskip 
from a field theory approach\footnote{This is based on the 
talk of D.\ Ghilencea at  the workshop of EC-Research Training Network, 
``The Quantum Structure of Spacetime and the Geometric Nature of
Fundamental Interactions'', 
 Physics Institute, Katholieke Universiteit Leuven, Leuven, 
13-19 September 2002. }
}
}
\vspace{0.4cm}
\bigskip 
\vspace{0.93cm} 

{\bf D.M.\ Ghilencea$^a$} and 
{\bf S.\ Groot Nibbelink$^{b,\dagger}$}
\\
\vspace{0.9cm} 
$^a${\it DAMTP, CMS, University of Cambridge} \\
{\it Wilberforce Road, Cambridge, CB3 0WA, United Kingdom.}\\
\bigskip 
$^b${\it Department  of Physics and Astronomy, University of Victoria,\\}  
{\it PO Box 3055 STN CSC, Victoria, BC, V8W 3P6 Canada,}\\
$^\dagger${\it CITA National Fellow.} 
\end{center}
\vspace{1cm}
\begin{center}
{\bf Abstract}\\
\end{center}
{\small 
An effective field theory approach is introduced to compute 
one-loop radiative corrections to the gauge couplings due to 
Kaluza-Klein  states associated with a two-torus compactification.
The results are compared with those of the string in the field 
theory ``limit'' $\alpha'\ra 0$. The whole $U$ and the leading
$T$ moduli dependence of the gauge dependent part of the string 
corrections to the gauge couplings can be recovered  using the effective 
field theory  approach.}

\newpage
\newpage\setcounter{page}{1}
\section{Introduction.}

String theory provides  a consistent framework for investigating 
the physics of very high energies in general,  and that of 
additional compact dimensions in particular. The physics of extra
dimensions can be studied with some success on effective field
theory grounds as well. Establishing an exact
 link of the string results  with those of the effective field theory is 
however a difficult task. The latter theory  may reproduce 
some results from string theory in the limiting case  of an infinite 
string scale $M_S$ (zero slope $\alpha' \propto 1/M_S^2$). 
It is this point of view that we adopt
in our effective field theory analysis below, while seeking the exact
relationship with string results for the problem of radiative
corrections to the gauge couplings. However, the results we present are 
also  relevant for models with ``large'' extra dimensions, without 
any reference to string theory.

Upon compactification, provided some conditions are fulfilled, 
remnant effects of the extra dimensions may be present and affect 
the low energy physics, such as the gauge couplings of the theory.
For our discussion below we consider the 
example of a  two-torus compactification. The model we 
address is  an  $\cN=1$ supersymmetric orbifold with one $\cN=2$
sub-sector (e.g. $\bZ_4$ orbifold). This two dimensional 
sub-sector (``bulk''), if  charged under the gauge group of the
theory, may bring one-loop corrections to the gauge couplings of the
theory.  At the string level these  will be due to  Kaluza-Klein and winding 
modes excitations with respect to the additional compact dimensions. 
These corrections are well-known in string case and were investigated
in \cite{Dixon:1990pc}, see also e.g. \cite{Mayr:1993mq}, \cite{Mayr:1995rx}.

An effective  field theory approach to investigating such corrections 
to the gauge couplings due to the two extra dimensions  is also possible. 
Such an approach can only account for the effects of  
Kaluza-Klein states: the winding states effects  cannot be 
described by a field theory approach. However, in the limit of a 
large compactification radius (compared to the ultraviolet length 
scale $1/\Lambda$), one would hope 
that the effects of winding states are minimised.  The reason for this 
would be that the above condition corresponds on the string side to 
the limit $\alpha'\ra 0$ in which case the mass of the winding states
is significantly larger than that of the  momentum modes. 
Given the {\it infinite} number of these states it is not actually  clear 
that they decouple in the low energy limit. Indeed, the 
density of momentum and winding modes is $\rho=R+1/R$, thus winding 
states may still affect the gauge couplings even if $R$ is  very 
large \cite{Kutasov:1990sv} in string units. Comparing the effective 
field theory result (due to effects of
 momentum modes only) with that of the string
(in its $\alpha'\!\ra \!0$ limit) will allow us understand whether 
the winding states  have any UV effect 
on the gauge couplings. Previous effective field theory approaches 
\cite{Taylor:1988vt} to such radiative corrections  were restricted to 
understanding the UV behaviour of the couplings in the presence of 
extra dimensions. The exact link with string theory was recently 
addressed  in \cite{Ghilencea:2002ff}.

We thus consider an $\cN=1$ orbifold with a $\cN=2$ sub-sector, and
attempt to  keep a general approach, without unnecessary model
dependence. To begin with we note that the radiative corrections 
to the ``bare'' or string coupling $\alpha_u$,
induced at one-loop level in the two-torus compactification 
are usually written  as 
\begin{equation}
\alpha _{i}^{-1}(Q) = \alpha _{u}^{-1}+\frac{b_{i}}{2\pi } \ln
\frac{M_{S}}{Q}+\Delta _{i} +\cdots,
\label{gauge}
\end{equation}
Here $Q$ is a low energy scale, 
$b_i$ the $\cN=1$ beta function 
coefficient, and $M_S$ is the ultraviolet scale
(string scale). 
The first logarithm in eq.\ (\ref{gauge}) accounts for the one loop
effects of the massless states of the model. These states can include 
for example ``twisted'' string states (candidates for the MSSM-like matter 
fields), and also gauge bosons' contributions. Computing their overall 
radiative effect on the gauge couplings requires an infrared (IR) and 
ultraviolet (UV) regularisation.
The second term $\Delta_i$ is due to massive Kaluza-Klein and, in 
string case winding states as well. The dots stand for higher order
corrections and for mixing effects between the massless and
$\cN=2$ massive states sectors.

\section{One-loop corrections from the string.}

We first review some details of the  string calculation 
for the  one-loop corrections $\Delta_i$ 
to the gauge couplings \cite{Dixon:1990pc}.
At string level there  exists an  additional, gauge independent 
(universal) correction to the gauge couplings, which was  
``absorbed''  into the (re)definition of the ``bare'' or string 
coupling $\alpha_u$. This ensures that at the string level 
this coupling  is invariant under the symmetry
$SL(2,\bZ)_T\times SL(2,\bZ)_U\times \bZ_2^{T\leftrightarrow U}$ 
of the (heterotic) string \cite{Nilles:1997vk}.

The separation of the radiative effects on the gauge couplings 
into massless and  massive modes contributions in (\ref{gauge}) 
in the string calculation is not imposed by a string symmetry
or principle.  
In general this is done because string calculations only compute 
the corrections due to {\it massive} modes  alone. The expression for 
$\Delta_i$  in string case (hereafter denoted $\Delta_i^H$) is 
\cite{Dixon:1990pc,Kaplunovsky:1987rp}
\begin{equation}
\Delta _{i}^H=\frac{\overline{b}_{i}}{4\pi }\int_{\Gamma }
\frac{d\tau_1 d\tau_2}{\tau _{2}}\left( Z_{torus}-1\right) 
\label{sss}
\end{equation}
Here  $\overline{b}_{i}$ is the beta function coefficient 
associated with the ${\cal N}=2$ sub-sector of the $\cN=1$ orbifold.
 $Z_{torus}$ is the two-torus 
string partition function.
The modulus $\tau =\tau _{1}+i\tau_{2}$ of the world sheet 
torus is integrated over the fundamental domain 
$\Gamma =\{\tau _{2}>0,|\tau _{1}|<1/2,|\tau |>1\}$. 
As the massless states contribution is added {\it separately} in 
eq.\ (\ref{gauge}), the massless (``zero'') mode contribution has 
been subtracted out in (\ref{sss}) and accounted for by ``-1''. One 
can show  that the integral over
$Z_{torus}$ alone is  modular invariant but is (infrared) divergent.
However, by subtracting the massless contribution the result is (made)
finite, but is not modular invariant anymore due to this last
contribution. Briefly, modular invariance does {\it not} ensure a
finite string result. Finally,  there seems to be  no reason to 
integrate this massless modes contribution ``-1'' over the 
fundamental domain $\Gamma$, except to  make the string result finite.
Quantitatively, $\Delta_i^H$ can be written as \cite{Dixon:1990pc}
 \begin{equation}
  \! \!
 \Delta _{i}^H\!=\!\frac{\overline{b}_{i}}{4\pi }\int_{\Gamma }\!
 \frac{d\tau}{\tau_{2}}
 \left[\frac{T_2}{\tau_2} \sum_{A} e^{-2\pi i T \det A}\,
 \exp\Bigl[
 {-\frac{\pi T_2}{\tau_2 U_2} 
 \Bigl\vert
 (1 ~ U) 
 A \mbox{\small$\Bigl(\!\! \begin{array}{c} \tau \\  1\end{array} \!\!\Bigl)$} 
 \Bigr\vert^2} \Bigr]
 -1\right];\quad
 A =\left(\!\!\! 
 \begin{array}{cc} n_1 & p_1\\ n_2 & p_2 \end{array}
 \!\!\!\right)\!\!\label{gag}
 \end{equation}
where  integers $p_{1,2}$ are (Poisson re-summed) Kaluza-Klein levels
and $n_{1,2}$ are winding modes. 
The moduli fields $T = T_1 + i T_2$,  $U = U_1 + i U_2$ can be expressed
in terms of the anti-symmetric tensor background $B_{ij} = B \ge_{ij}$
and the radii $R_{1,2}$ and angle $\theta$ of the two-torus as 
$T=2 [ B + i {R_1 R_2 \sin \gth}/{(2 \ga')}]$,
$U = {R_2}/{R_1} \exp{(i \gth)}$.

The integral in eq.(\ref{gag}) can be written as a sum over the ``orbits'' of
the modular group $SL(2,\bZ)$. This is just a sum over classes of
matrices $A$ with entries integer numbers  labelling (Poisson re-summed) 
Kaluza-Klein and winding levels, giving  \cite{Dixon:1990pc} 
\begin{equation}\label{sum}
\Delta_i^H=\frac{{\overline {b}}_i}{4 \pi}\left[
\cJ ^{(A=0)}+\cJ ^{(\det A=0)}+\cJ ^{(\det A\not=0)} +\int_\Gamma
\frac{d\tau_{1}d\tau _{2}}{\tau _{2}} (-1) \right]_{reg}
\end{equation}
The subscript ``reg'' indicates that the splitting
of the integrals in  eqs.(\ref{sss}), (\ref{sum})  only makes sense in the 
presence of  an infrared regulator. For example, each of the separate 
terms can be  multiplied by $\cR(\tau_2)=(1-\exp(-N/\tau_2))$ with
$N\ra\infty$. This regularisation breaks modular invariance, but 
other IR regularisation schemes, which are 
modular invariant \cite{Kiritsis:1994yv} can be used.

The three (regularised) contributions in eq.(\ref{sum})  are 
classified in function of the matrices $A$ and correspond to: 
the zero orbit $\cJ^{(A=0)}$,  the degenerate orbit $\cJ^{(\det
A=0)}$, and the non-degenerate orbit $\cJ^{(\det A\not=0)}$. 
$\cJ^{(A=0)}$  is due to infinitely many original  Kaluza-Klein modes.
 $\cJ^{(\det A=0)}$ is due to a mixing of momentum and winding
modes (if $n_1=p_1=0$), or momentum modes alone (if
$n_{1,2}=0$) or winding modes alone (if $p_{1,2}=0$).
In the limit of an infinite string scale or $\alpha'\ra 0$ the
momentum modes contribution is dominant.
 $\cJ^{(\det A\not=0)}$ is due to a mixing of momentum and winding
modes and is vanishing in the limit of an infinite
string scale or  $\alpha'\ra 0$ when winding modes' effects are
suppressed. Quantitatively, one has
\begin{eqnarray}
\cJ ^{(A=0)}&=& \int_{\Gamma } \frac{d\tau_{1}d\tau _{2}}{\tau _{2}^2}\, T_2=
\frac{\pi}{3} T_2, \label{piover3}\\
\cJ ^{(\det A=0)}&=& \ln N -  
\ln \Bigl[  4 \pi  e^{-2\gamma_E} T_2 \, U_2 \,\, | \get(U) |^4
\,\,\Bigr]+\cO \left( ({T_2
U_2}/{N})^{\frac{1}{2}}\right)\label{detA=0}\\
\cJ ^{(\det A\not=0)}&=&
-\ln \prod_{n_1=1}^{\infty} \left| 1-e^{2 \pi i \,n_1\, T }\right|^4,\
\qquad  \cJ ^{(\det A\not=0)} \ra 0 \textrm{   if   } \alpha'\ra 0
\label{windingmode}\\
\int_{\Gamma }
\frac{d\tau_{1}d\tau _{2}}{\tau _{2}} (-1)\bigg\vert_{reg.}
&=& -\ln N +\ln{3\sqrt 3} e^{-1-\gamma}/{2},\label{massless}
\end{eqnarray}
The term $\cO((T_2 U_2/N)^{1/2})$ is just a correction depending on
the infrared regulator and vanishes in the limit of removing it,
$N\ra \infty$.
Adding together the above equations, one obtains the final result
\begin{equation}\label{t2formp}
\Delta _{i}^H =-\frac{\overline{b}_{i}}{4\pi }\ln \left\{ 
C_{reg}\,U_2\,{|}\eta (U){|}^{4}\, 
T_{2}\left| \eta \left( {T}\right) \right|^{4}\right\}
+\cO \big( ({T_2 U_2}/{N})^{\frac{1}{2}}\big),
\end{equation}
where $C_{reg}=8\pi e^{1-\gamma}/(3\sqrt{3})$ is a regularisation scheme
dependent constant (for the DR scheme  \cite{Foerger:1998kw}, 
$C_{reg}=4\pi e^{-\gamma}$).

\section{One-loop corrections from a field theory approach.}

A field theory approach to computing the effects of 
the two extra dimensions of the two-torus on the gauge couplings 
can only sum the effects due to Kaluza-Klein states (no winding modes).
The calculation we present is also relevant in the context of 
models with ``large'' compactification radii  (relative to the UV cut-off 
length) and  without any reference to the string.
As the massless states' effects are already accounted for in 
eq.(\ref{gauge})  the one-loop
radiative corrections to the gauge couplings due to massive modes 
alone is (hereafter denoted  $\Delta^*_i$)
\begin{equation}\label{QFTthresholds1}
\Delta^*_{i} = \frac{1}{4\pi}\sum_{i} T(R_i) 
\sum_{m_{1,2} \in \bZ}' \int_{\xi}^{\infty}\frac{dt}{t} 
\, e^{-\pi\,t M_{m_1,m_2}^2/\mu^2 },
\end{equation}
The result of summing a finite or infinite tower of Kaluza-Klein 
states to the gauge couplings is divergent. Since the integral 
above is divergent in the UV ($t\ra 0$), we introduced a 
UV dimensionless (proper time) regulator 
$\xi\ra 0$ as the lower limit of the integral, and an arbitrary
(finite) mass scale $\mu$. A ``prime'' on the sum indicates that  it
does not include the effects of the massless state $(m_1,m_2)\not
=(0,0)$, which is already present in (\ref{gauge}).
Finally, the mass of the Kaluza-Klein states is \cite{Dienes:2001wu}
\begin{equation}\label{kkmass}
M^2_{m_1,m_2}=
\frac{1}{\sin^2\theta}\left[\,\frac{m_1^2}{R_1^2}+\frac{m_2^2}{R_2^2}-
\frac{2 m_1 m_2 \cos\theta}{R_1 R_2}\,\right]
\,\, =\,\, \frac{\vert m_2-U m_1\vert^2 }{ (\mu^{-2} T_2(\mu)) U_2},
\end{equation}
where 
\(
{T} (\mu) \equiv i T_2 (\mu)= i \mu^2 R_1 R_2 \sin\theta
\)
and the complex moduli $U$ is identical to that in the string case, 
$U=R_2/R_1 \exp(i\theta)$. For the particular case of an orthogonal torus 
$\theta=\pi/2$ one recovers a more familiar
mass formula: $M_{m_1,m_2}^2 = m_1^2/R_1^2+m_2^2/R_2^2$,
and if the radii are equal one obtains that 
$M_{m_1,m_2}^2 =(m_1^2+m_2^2)/R_2^2$.

The calculation of the integral for $\Delta^*_i$ is rather technical, so 
we only quote the  final result \cite{Ghilencea:2002ff}  
\begin{eqnarray}\label{result}
\Delta^*_i&=& - \frac{{\overline {b}}_i}{4 \pi} 
\ln\Big[4 \pi e^{-\gamma_E} \, e^{-{T_2^*}} \, {T_2^*} \, U_2
\, \vert \eta(U)\vert^4\Big], \\
\nonumber\\
T_2^*&\equiv& \Lambda^2  R_1 R_2 \sin\theta,\qquad
\Lambda^2\equiv {\mu^2}/{\xi},\qquad
 \max\left\{{1}/{(R_2 \sin\theta)}, {1}/{R_1}\right\} \ll \Lambda
\label{rs}\end{eqnarray}
where $\overline b_i$ is a sum over the beta function contributions of
those states with a  Kaluza-Klein tower 
and $\Lambda$ is the UV scale 
associated with the UV regulator $\xi$.
This result  for $\Delta^*_i$   holds true only in the limit of ``removing''
the dependence on the regulator $\xi\ra 0$, when additional 
corrections in $\xi$ vanish.  This condition is converted into 
bounds  on the mass scales of the theory presented in (\ref{rs}).
Notice that $R_2 \sin\theta$ plays the role of an effective radius.

\section{A comparison with string theory corrections.}

We can now  compare the field theory result  (\ref{result}) with 
the heterotic string expression  (\ref{t2formp})
for the gauge thresholds. For this comparison to make sense we
note that one should actually compare the field theory results with
the limit $\alpha'\ra 0$ of the string results (\ref{t2formp}), when 
the effects of the winding modes, not included in the  field theory
calculation,  are suppressed.

We observe that the $U$ dependence of  the two equations
is identical. As both the effective field theory and the string theory
mass spectra are $SL(2,\bZ)_U$ symmetric, and $U$ is by definition
$\alpha'$ independent (i.e. scale independent), it may not be 
surprising that the $U$ dependence of the final string result can 
be entirely  recovered on  effective field theory grounds.

Regarding the $T$ dependent part  of the two results (\ref{t2formp}), 
(\ref{result}) we note the following. The field theory UV cut-off in
(\ref{result}) can be identified 
with the string scale:  $\Lambda^2 \equiv \mu^2/\xi\ra 1/\alpha'$. 
Removing the regulator, $\xi\ra 0$  on the field theory side in 
(\ref{QFTthresholds1}) 
corresponds to an infinite string scale or $\alpha'\ra 0$. 
In this limit eq.\ (\ref{t2formp}) (see also (\ref{piover3}), (\ref{detA=0})) 
has  quadratic and logarithmic divergences in the string scale
similar to those in  eq.(\ref{result}).  
However, the coefficient of the quadratic divergence 
is different: $T_2^*$ in $\Delta^*_i$, and $(\pi/3)\, T_2$ in
$\Delta^H_i$. In the field theory case this coefficient is
regulator {\it dependent}, thus it can be {\it chosen} such that this 
difference is avoided. Such choice  provides the appropriate
definition for the UV cut-off scale of an effective field theory, 
compatible with  the modular invariance of its string embedding.
We conclude that the full UV structure of the string thresholds 
in the limit $\ga' \ra 0$ can be obtained on pure field theory
grounds, except that the coefficient of the quadratically divergent
part is arbitrary in the field theory approach. 

To explain how the factor $\gp/3$ arises in the heterotic string, 
and its link with  winding modes effects, one should analyse
its origin in eq.\ (\ref{piover3}).  This factor  is 
consequence of the integration over the fundamental domain $\Gamma$, 
and is thus an effect of modular invariance symmetry and indirectly, 
of the winding modes. As this symmetry is not present in the effective 
field theory, the factor $\pi/3$ cannot be recovered. Thus winding modes
indirectly control the coefficient of the term quadratic in string scale
($T_2 \propto M_S^2$), therefore they  do have an UV role, even in the
limit $\alpha'\!\ra\! 0$.

We would like to end with an additional  remark on the field theory
limit $\alpha'\ra 0$ of the string result, eq.\ (\ref{t2formp}). 
A closer look at the final (infrared regularised) 
string result of eq.\ (\ref{t2formp})
shows that higher corrections due to the infrared regularisation 
$\cO((U_2 T_2/N)^{1/2})$ are discarded in the final result for $\Delta_i^H$
when $N\ra  \infty$.  This is true if $T_2$ is finite. However, 
in the field theory limit of $\alpha'\ra 0$ or equivalently
$T_2\ra\infty$ (with $T_2$ expressed in string units) 
the term $\cO((U_2 T_2/N)^{1/2})$ is not   vanishing. 
Therefore this term  should be kept  in the field theory limit of 
the string result $\Delta_i^H$.
This issue requires an investigation at the string 
level, to clarify if any string symmetry left after compactification
can still impose the order to take the limits of removing the 
infrared regulator $N\ra \infty$ and the field theory limit 
$T_2\ra \infty$ on the string result. This is relevant because these
two limits do not commute. For further discussions on 
this and its relationship to the field theory
approach see recent results  in \cite{IRUV}.

\section{Conclusions.}

We provided an  effective field theory approach to computing 
the corrections to the gauge couplings due to Kaluza-Klein states 
and a comparison with the string result which also includes 
the effects of the winding states.

While the heterotic string calculation is well known  and studied 
in the literature, we attempted to emphasize some points which are 
usually overlooked: one is that that the finiteness of the string
threshold correction $\Delta_i^H$ is not due to modular invariance. 
 It is the massless modes' contribution
which is introduced to ``make'' the result finite, by 
subtracting the (infrared) divergence
due to the massive momentum and winding modes. The integration of the 
massless modes contribution (non-modular invariant) over the
fundamental domain remains a procedure to make the string result 
finite, but is not required or supported by a string symmetry. 
One should clarify at the string level if any string symmetry
can dictate the order to take the limits of removing the
string infrared regulator and the field theory limit, since these
two limits do not  commute.

The effective field theory approach to computing 
corrections to the gauge couplings due to the  two extra dimensions 
of the torus provides results similar to those of the string in 
the limit of an infinite string scale. The full 
dependence on the shape moduli ($U$) 
is recovered. In addition, the UV divergences of the
string result  in the limit $\alpha'\ra 0$ were reproduced by
the field theory calculation.  The coefficient in front of
the leading UV divergences in  string theory  (in the limit
$\alpha'\ra 0$) turns out to be equal to $\pi/3$ as a result 
of modular invariance symmetry,
while at field theory level this coefficient is regularisation
scheme dependent.  

\noindent
\section*{Acknowledgements.}

\noindent
 This work was supported  by PPARC UK (D.G.), 
 and in part by CITA and NSERC (S.G.N.).

\end{document}